\documentclass[acmsmall,screen=True]{acmart}\usepackage[]{graphicx}\usepackage[]{color}
\makeatletter
\def\maxwidth{ %
  \ifdim\Gin@nat@width>\linewidth
    \linewidth
  \else
    \Gin@nat@width
  \fi
}
\makeatother

\definecolor{fgcolor}{rgb}{0.345, 0.345, 0.345}

\usepackage{framed}
\makeatletter
 {\par\unskip\endMakeFramed%
 \at@end@of@kframe}
\makeatother

\definecolor{shadecolor}{rgb}{.97, .97, .97}
\definecolor{messagecolor}{rgb}{0, 0, 0}
\definecolor{warningcolor}{rgb}{1, 0, 1}
\definecolor{errorcolor}{rgb}{1, 0, 0}
\usepackage{alltt}

\AtBeginDocument{%
  \providecommand\BibTeX{{%
    \normalfont B\kern-0.5em{\scshape i\kern-0.25em b}\kern-0.8em\TeX}}}

\setcopyright{acmcopyright}
\copyrightyear{}
\acmYear{}
\acmDOI{}

\acmConference[CSCW '20]{CSCW '20: Conference on Computer-Supported Cooperative Work and Social Computing}{October 17--21, 2020}{Minneapolis, MN}
\acmPrice{}
\acmISBN{}

\usepackage[]{natbib}

\IfFileExists{upquote.sty}{\usepackage{upquote}}{}
\begin{document}

\title[How individual behaviors drive inequality in online community sizes]{How individual behaviors drive inequality in online community sizes: an agent-based simulation}

\author{Jeremy Foote}
\email{jdfoote@purdue.edu}
\orcid{0000-0002-5331-1916}
\affiliation{%
  \institution{Purdue University}
}

\author{Nathan TeBlunthuis}
\email{nathante@uw.edu}
\orcid{0000-0002-3333-5013}
\affiliation{%
  \institution{University of Washington}
}

\author{Benjamin Mako Hill}
\email{makohill@uw.edu}
\orcid{0000-0001-8588-7429}
\affiliation{%
  \institution{University of Washington}
}

\author{Aaron Shaw}
\email{aaronshaw@northwestern.edu}
\orcid{0000-0003-4330-957X}
\affiliation{%
    \institution{Northwestern University}
}

\renewcommand{\shortauthors}{Foote et al.}

\newcommand{\mywidth}{.6\columnwidth}
\newcommand{\appwidth}{.48\columnwidth}

\begin{abstract}

Why are online community sizes so extremely unequal? Most answers to this question have pointed to general mathematical processes drawn from physics like cumulative advantage. These explanations provide little insight into specific social dynamics or decisions that individuals make when joining and leaving communities. In addition, explanations in terms of cumulative advantage do not draw from the enormous body of social computing research that studies individual behavior. Our work bridges this divide by testing whether two influential social mechanisms used to explain community joining can also explain the distribution of community sizes. Using agent-based simulations, we evaluate how well individual-level processes of social exposure and decisions based on individual expected benefits reproduce empirical community size data from Reddit. Our simulations contribute to social computing theory by providing evidence that both processes together---but neither alone---generate realistic distributions of community sizes. Our results also illustrate the potential value of agent-based simulation to online community researchers to both evaluate and bridge individual and group-level theories.
\end{abstract}

% ACM Classfication





\maketitle

\section{Introduction}

Massive inequality in membership size is a feature of virtually every platform hosting online communities. A few very large communities attract the vast majority of contributors while most communities receive almost none. This pattern occurs in peer production communities like open source software and wikis as well as discussion-oriented communities like forums. For example, in January 2017 the ``r/AskReddit'' community on Reddit had over 680,000 unique contributors who made over 5.3 million comments. The next most active community on Reddit, ``r/politics'' had around 156,000 contributors and 2.2 million comments. Meanwhile, the median number of contributors to a Reddit community that month was three and the median number of total comments was five.\footnote{All data from Reddit (\url{https://reddit.com}) used in this paper was gathered and republished by Pushshift (\url{https://pushshift.io}).}

Why are some online communities so much larger than others? Why do people join online communities? Most prior answers to the first question in social computing research treat group size as the result of mathematical mechanisms drawn from physics such as cumulative advantage or preferential attachment. These approaches minimize agency and lack a clear link to individual-level behavior. The body of work answering the second question has provided explanations of how people make these individual decisions but only in the context of a single community. 
Social computing research suggests three sequential subprocesses that influence individual decisions to join or leave communities. In the first step, people learn about communities through social ties and social influence. We call this \textit{social exposure}. In a second step following exposure, people decide whether to participate for the first time or not (i.e., to join a community). Finally, after a period of activity, people decide to discontinue their participation. Prior work treats joining and exit decisions as a function of individuals' motivations, abilities, and/or expectations that a given community will satisfy their goals in some way. We refer to this approach to joining and exit decisions as \textit{individual expected benefits} (IEB). 

These individual-level explanations typically do not attempt to explain macro-level phenomena like the distribution of participants across communities. Logically, group sizes must emerge as a function of individual decisions about which groups to join. But how well do individual joining processes explain macro outcomes like membership size? 
Understanding higher-level implications of individual-level behaviors is often difficult. In social systems like online communities, individual behaviors are interdependent and combine to produce macro-level patterns in complicated ways that are hard to predict. This ``micro-macro divide'' makes it very difficult to directly test findings across levels.

One approach to bridging this divide is agent-based simulation (or ABS, also known as agent-based modeling). ABS involves building empirically-informed formal approximations of theories of individual behavior, using computers to step through simulated interactions between individual agents behaving according to the formalized theories, and comparing the macro-level outcomes of these simulations against empirically-observed data.

To bridge the micro-macro divide in online community research, we formalize models of theories of social exposure and IEB decisions and use ABS to simulate computational agents acting according to these models across a range of simulated situations. We also simulate a novel joint model of social exposure and IEB. We evaluate our models and the underlying theories by assessing how well agents reproduce the patterns of community size observed empirically in Reddit. We find that simulated agents acting according either social exposure or IEB alone do not produce empirically plausible community size distributions. In contrast, when the two sub-processes of exposure and IEB decisions act in tandem, sufficient positive feedback emerges to produce highly skewed distributions similar to that of Reddit.

This paper makes three primary contributions. First, we provide a framework for connecting micro-level social computing research influenced by social psychology with macro-level research on organizational behavior and group and population dynamics. In doing so, we show how higher-level patterns can enrich our understanding of individual behavior, and vice versa. Second, we provide arguments and evidence for the usefulness of agent-based simulation in social computing research. Finally, we provide a theoretical synthesis between social exposure and IEB decisions and show that both together provide a good explanation for the extreme inequality in online community sizes.

\section{Background}

Social computing research on community participation and community growth is divided into two largely distinct bodies of micro- and macro-level scholarship. To motivate our use of agent-based simulation as a bridge between the two levels, we briefly review salient examples of each. First, we consider micro-level explanations of community membership that focus on two sub-processes: how people learn about communities through social exposure and how they decide to join or leave them based on expected benefits. We then discuss prior explanations of community size distributions from a macro-level perspective, underscoring that such explanations typically have little to say about individual-level behaviors. 
Next, we introduce ABS as a method for evaluating the impact of micro-level community joining behavior on the macro-level distribution of community sizes.
Finally, we discuss how micro-level processes may contribute to macro-level behavioral outcomes and why macro-level consequences matter for micro-level models.

\subsection{Explaining individual-level community joining and exit}
\subsubsection{Social exposure}

While people may learn about new communities in multiple ways, exposure often occurs through social ties \citep{kraut_building_2012}. Social exposure-based explanations of online community joining are one example of how social ties predict the adoption of computing behaviors. For example, the more friends who adopt a technology product \citep{bhatt_predicting_2010}, patterns of behavior \citep{bakshy_social_2009, state_diffusion_2015}, or shared information \citep{bakshy_role_2012}, the more likely a focal person is to do the same. A few studies have looked explicitly at the decision to join online groups and have found the same dynamics: the more friends a person has in a group, the more likely they are to join it \citep{backstrom_group_2006, kairam_life_2012}.
Social exposure to new online communities is a common feature of participation on the Internet and users of online communities are frequently being exposed to new communities. For example, in January 2017, 792,643 comments on Reddit---over 1\% of all public comments made on Reddit that month---included links to other Reddit communities.

\subsubsection{Decisions based on individual expected benefits}

Once exposed to a set of communities, people join online communities in order to advance their goals and to meet their needs \citep{klandermans_why_2004, resnick_starting_2012}. Individuals decide whether and how to participate in a community based on their personal attributes, motivations, and experiences. Potential members reason prospectively about the community and whether they can imagine themselves as a part of it \cite{antin_readers_2010, antin_my_2011, antin_technology-mediated_2012}.  Over time, individuals will remain or leave depending on how existing members respond to them (e.g., do they bite the newbies or reach out to offer support?) or based on shifting perceptions of the community, other members, or their own role \citep{bryant_becoming_2005, halfaker_dont_2011, halfaker_rise_2013, morgan_evaluating_2018, panciera_wikipedians_2009}. As part of these decisions, people may estimate the impact and importance of their contributions. For example, when the government of the People's Republic of China eliminated access to Wikipedia for their citizens, thereby decreasing the editor community and audience of Chinese Wikipedia,  Chinese-speaking Wikipedians from the rest of world dramatically reduced their participation \citep{zhang_group_2011}. All else equal, larger communities with larger audiences tend to attract more participants.
After joining a group, some people deepen their commitment to it and develop attachments in the form of identity and/or social bonds \citep{kraut_building_2012}. The extent to which a person has these attachments also influences whether they continue to participate or leave \citep{danescu-niculescu-mizil_no_2013, kairam_life_2012}. Eventually, individual community members reach a peak of participation and exit a community when it no longer helps to meet their goals.

To summarize this prior work in a schematic way, people decide to join and participate in online communities as long as they expect that the benefits they receive outweigh the costs of participating. 
We use the term \textit{individual expected benefits} (IEB) to describe this way of thinking about joining decisions. The IEB approach is illustrated in the final chapter of the book \textit{Building Successful Online Communities} \citep{kraut_building_2012}, wherein \citet{resnick_starting_2012} provide an equation-based model intended to summarize how people weigh costs and benefits when deciding whether to join an online community. Like the rest of the book, the model is presented as part of extensive literature review and attempts to integrate and summarize a rich body of work on the topic of community joining and exit decisions in social computing theory.

In \citeauthor{resnick_starting_2012}'s model, benefits fall into two categories: participation benefits and early adopter benefits. Participation benefits are a function of the eventual size of the community and might include social relationships, entertainment, and information. Early adopter benefits are benefits which only accrue to early members and include influence, reputation, or even revenue sharing. The costs to join a community (which they call ``startup costs'') include learning new software, learning the norms of a new community, and building reputation and social ties. If someone believes that the expected utility of the participation benefits plus the early adopter benefits outweigh the costs then they will join that community or continue to participate. If not, they leave the community or never participate in the first place.
% \footnote{This approach echoes \citet{olson_logic_1965}'s emphasis on ``selective incentives'' for individuals to engage in collective action of any kind.}

\subsection{Explaining inequality in online community size}

\begin{figure}
    \centering
    \includegraphics[width = \mywidth]{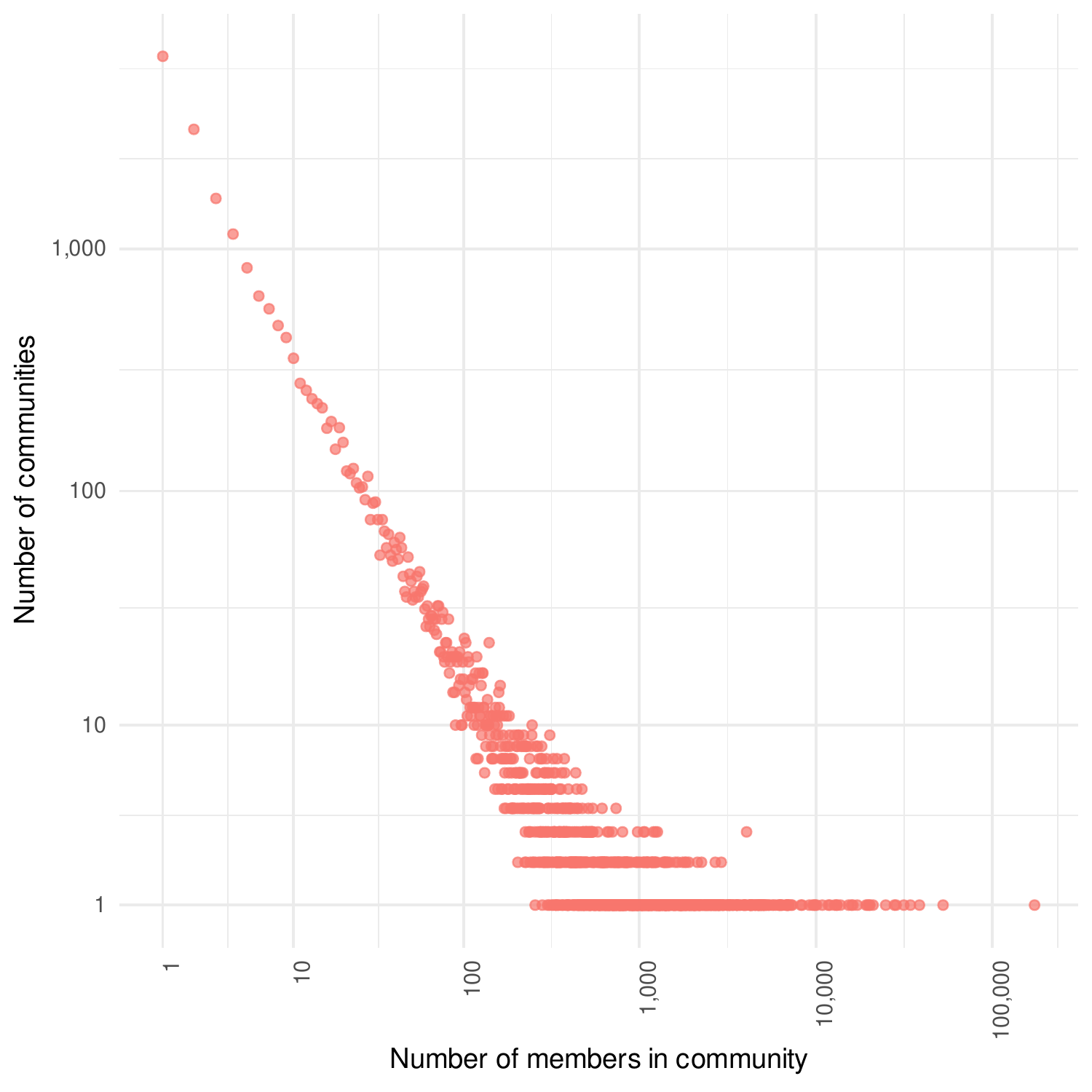}
    \caption{Distribution of members per subreddit in January 2017. The x-axis is the number of users with at least five comments in that subreddit in the month. The y-axis is the count of communities with that number of users. Both axes are log-scaled.}
    \label{fig:subreddit_sizes}
\end{figure}

Although distributions of community size emerge as a consequence of individuals joining and leaving communities, prior work has analyzed these distributions without making serious efforts to explain their relationship to individual-level behaviors. One clear takeaway from studies of community size across a range of platforms and contexts is that distributions of community sizes are highly skewed \citep{johnson_emergence_2014}. For example, in one month on the platform Reddit, twelve subreddits (i.e., topic-based communities) had over 100,000 unique contributors, while over 44,000 subreddits had fewer than five contributors (Figure \ref{fig:subreddit_sizes}). Free/open source software communities \citep{graves_open_2013, healy_ecology_2003, crowston_free/libre_2008}, wikis and other peer production production projects \citep{benkler_peer_2015}, online discussion forums \citep{jones_empirical_2002, panek_effects_2018}, and many other platforms follow similar patterns.

Whether identified as following a scale-free ``power law'' \citep{adamic_power-law_2000, barabasi_emergence_1999}, ``weakly'' scale-free, or merely log-normal \cite{broido_scale-free_2019} distributions, explanations of community size typically invoke cumulative advantage and/or preferential attachment \citep{barabasi_emergence_1999, merton_matthew_1968}. In these models, current community size is the only variable used to predict future community size. In short, popular communities become ever more popular. Drawing from models used in physics, these explanations emphasize mathematics of accumulation without providing credible social mechanisms of individual action that might act as reasonable approximations of the mathematics. Sociologists have referred to these types of explanations as ``undersocialized'' accounts \citep[e.g.,][]{frank_strategic_1993}.

Could models of social exposure and IEB act as social mechanisms which reproduce the broader dynamics of cumulative advantage or preferential attachment, and thus explain higher-level patterns of inequality in community sizes based on empirically-grounded, individual-level research? In that social computing scholars studying joining processes have focused on individual-level and group-level outcomes, they have not attempted to understand patterns in populations of groups. As a result, we simply do not know.

Of course, researchers have considered why groups grow or founder. Most of this prior work also focuses on just one level. For example, some studies explain variation in group outcomes based on group-level behavior. In this work communities appear as independent entities and features of members, their interactions, or institutions predict outcomes like size or longevity \citep{crowston_social_2005, crowston_core_2006, kittur_beyond_2010, kraut_role_2014, shaw_laboratories_2014}. Although valuable, this work tells us little about the distribution of community sizes because group membership is presumed and differences in outcomes arise from things that happen within communities. 

Other studies analyze relationships between groups---e.g., whether overlaps in group membership predict group survival or growth \cite{teblunthuis_density_2017, wang_impact_2013, zhu_selecting_2014, zhu_impact_2014}. Again, this research explains variation between community outcomes and does not account for the overall distributions of community size.

\subsection{Agent-based simulation}

In short, previous research provides little insight into how well individual behavioral models of joining and exit explain inequality in online community sizes. One explanation for this gap in our understanding is the challenges of drawing inferences across levels of theorizing and analysis. It is difficult to test competing theories of joining and community size simultaneously because doing so requires studying individual-level processes across many organizations and communities. This poses a barrier to research because many challenges involved in conducting empirical social computing research scale with the number of communities and contexts involved. For example, issues of access to data on users and their motivation, challenges related to both the size and diversity of data, and nuts-and-bolts issues related to the navigation of idiosyncratic features of multiple communities make it challenging enough for researchers to study detailed user-level behavior within a single community. Doing it across thousands or even millions of communities is often simply not possible. 

Prior work in the social sciences describes this type of challenge in terms of a ``micro-macro divide'' \cite{opp_modeling_2011}. To address these challenges, researchers in ecology, economics, and sociology have characterized ``agent-based complex systems'' \citep{grimm_pattern-oriented_2005} where properties of larger systems emerge from the decisions of interdependent agents. These disciplines often employ agent-based simulations in order to bridge the micro-macro divide \citep{grimm_pattern-oriented_2005, wilensky_introduction_2015}. Agent-based simulations (ABSs) capture important aspects of a context in the form of simple rules that agents follow. Researches then create virtual experiments by modifying these rules in order to identify the conditions under which interacting agents produce various higher-level outcomes \citep{lazer_network_2007}. At their best, such models reveal patterns of emergent collective behavior that even thoughtful analysis of micro-level action might never predict (and vice-versa). In one of the earliest agent-based simulations, Schelling \citep{schelling_dynamic_1971} showed how a world in which agents hold even a very weak preference to live near someone who resembles them becomes completely segregated---without any reference to geography, jobs, moving costs, social networks, and almost everything that real people seem to consider in deciding to move. 

Theories like critical mass theory \citep{marwell_critical_1993} and complex contagion \citep{centola_complex_2007} were developed with the help of agent-based simulations and have been influential in social computing research \citep{raban_empirical_2010, romero_differences_2011}.  Despite this history of benefiting from ABS, a strong fit with social computing research questions, and calls for use from prominent scholars \citep{ren_agent_2014}, agent-based simulations remain extremely rare in HCI and social computing. 

Because they bypass many of the challenges with large-scale empirical research described above, ABSs offer a feasible approach to bridge theories and findings at different levels of analysis. They can allow theory and data at higher levels to influence micro-level mechanisms and provide grounded, validated explanations for macro-level patterns. ABSs also let us explore how variation in assumptions and models of individual behavior combine and aggregate to group and population dynamics and can thus help enrich theories of individual and group behavior.

\subsection{Bridging the ``micro-macro divide'' between joining/exit and community size}

An ABS approach allows us to explore the community size distributions that emerge when agents join and leave communities according to models of social exposure and/or IEB decisions. This approach helps to formalize questions and theories, test how changes to assumptions and parameters influence outcomes, and evaluate the simulated outcomes against empirical baselines.

Consider the implications of social exposure dynamics on community size distributions. Social exposure provides a plausible micro-level mechanism of cumulative advantage. In that larger communities have more members by definition, they also have more people who can talk about them. This could lead to more people learning about and joining larger communities than smaller ones. This would result in even more people talking about the larger communities in the future. The result is a positive feedback mechanism where success begets success \citep{van_de_rijt_field_2014}.

Joining decisions based on IEB could also contribute to skewed outcomes. If individuals take participation benefits and early-adopter benefits as described by \citet{resnick_starting_2012} into account as they decide which communities to join, they will gravitate towards those that are already large or that appear positioned to become largest. This should also create a feedback loop, where individuals join quickly growing communities, causing others to be even more likely to see them as quickly growing. As with social exposure, these dynamics could produce skewed group sizes.

Together, the positive feedback patterns created by social exposure and IEB decisions should amplify each other. With both forces operating jointly, people are both more likely to be exposed to larger communities as well as more likely to join them. Subsequently, these new members will expose others to these larger communities, who will also be more likely to join them, and so on. We would expect this to result in even more skewed distributions of community size than either model operating alone. 

\section{Methods}

We created a set of agent-based simulations to explore whether computational models of social exposure and/or IEB decisions generate distributions of community size that resemble those observed empirically. To do so, we formalized theories of social exposure and IEB decisions into algorithms that simulated agents that decide which communities to join or leave and which to share with others. For each of the micro-level mechanisms, we simulated a series of decisions and interactions among a population of agents over time to observe the macro-level consequences in a population of hypothetical communities. We show the results of these simulations and compare the distribution of hypothetical community sizes to a sample drawn from Reddit. Additional details about the general approach, how we generated each set of simulations, and the process we follow for comparing the simulated results against the empirical baseline appear below.

\subsection{Simulated models of community joining and exit}

We performed four families of simulations. First, a set of null models with agents joining and leaving communities randomly to provide a starting point for comparison. We then ran models which test the influence of social exposure and IEB decisions separately. Finally, we simulated a model that includes both social exposure and IEB decisions.

For each of the four sets of simulations, we parameterize key conceptual variables involved in community joining and exit and run each simulation with slightly varied parameter values. Doing so serves several purposes. First, the varied parameters help ensure that the results from a set of simulations represent a general, meaningful pattern rather than an artifact of any specific parameter. This is especially important because many of the parameter values are choices informed by prior research rather than precise empirical estimates. Second, by observing the results over a range of plausible values, we can evaluate variation in outcomes over different realizations of the theories we test. Finally, this approach reduces the risk that our findings reflect flukes that only appear in a limited set of conditions.

All of the models are intended to simulate populations of people and communities over time. Each simulation proceeds through a series of steps. In each time step, each agent sees a subset of communities (the exposure set). Agents then decide whether to leave any of their current communities and whether to join any of the communities in the exposure set. Our model stores the results of these decisions and then moves to the next agent until all agents have completed their decisions. The process iterates repeatedly over a series of time steps until we declare a stopping point.

We fix the size of each of our simulations to 9,000 agents and 200 communities. We use 9,000 agents because it is few enough to be computationally tractable but large enough to allow for complex dynamics to emerge. We chose 200 communities because this sets the ratio of members and communities close to what is observed empirically on Reddit. In January 2017, there were 3,578,907 active commenters on Reddit who commented in 78,201 subreddits, a ratio of approximately 90:2. 
We run each simulation for 24 time steps in order to allow the community size distributions to reach a steady state. Finally, we measure the number of members per community.

\subsubsection{Null models}

We first simulate a null model, in which exposure and participation are random. The null model does the following during each simulated time step: Each agent randomly leaves any community they currently belong to with probability $p_l$. Each agent draws a random sample from the pool of all communities with with probability $p_e$, and each agent randomly chooses to join each sampled community with probability $p_j$. 

We can observe how often people \textit{leave} online communities, and so we set a value for $p_l$ based on the proportion of people who leave subreddits each month on reddit. Of all user accounts leaving at least 5 comments in any given subreddit in January 2017, 56\% did not comment at least five times in that same subreddit in February. As a result, we set $p_l$ to $.56$. 
Exposure is less visible. People are exposed to new online communities but we don't know how many or which ones. Neither can we estimate the probability of participating after exposure. Because we assume that both are fairly low, we vary $p_e$ and $p_j$ over levels of $\{.01, .05, .1, .2\}$ for our null simulations.

\subsubsection{Social exposure models}

Theories of social exposure suggest that people learn about new communities through people they are already connected to. Our social exposure models capture this dynamic by randomly sampling $k$ community members from each of the focal agent's $n$ communities. Larger communities have more people and therefore more opportunities for social connections, so we set $k_i = ceil(log(|n_i| + 1))$ where $|n_i|$ is the current size of the $i$th community and $k_i$ is the number of ``neighbors'' sampled from that community. The $ceil$ function rounds up to the nearest integer, ensuring that we always sample at least one neighboring agent. 

During each time step, each of these $k$ neighboring community members samples up to $m$ of their other communities to share with the focal agent. We vary $m$ from $1$ to $4$ across different simulations. This process defines an exposure set for each agent of maximum size  $n * k * m$. The size of the exposure set can be smaller if the chosen fellow members belong to fewer than $m$ other communities or if they share a community already in the exposure set. An agent not yet belonging to any communities is exposed to a random sample of all of communities with uniform probability $p_e = .1$. Because social exposure theories have nothing to say about community exit, agents exit communities randomly with probability $p_l = .56$ as in our null models. From our null models, we found that choices for $p_e$ and $p_j$ were of little importance to the qualitative distributional properties of interest. As a result, we fix $p_e = p_j = .1$ and include models where $p_e = p_j = .05$ in the appendix as robustness checks.

No research that we know of provides insight into \textit{which} communities people are likely to share with others. We show one model designed to capture this ignorance in which agents simply choose randomly which communities to share. However, one argument of this paper is that individual-level theories can and should be informed by higher-level phenomena. If we consider the skew toward membership in large communities observed empirically, we might guess that agents are more likely to share large communities. We therefore also simulate a social exposure model where agents share the largest communities to which they belong.

\subsubsection{IEB decision models}

In order to formalize and simulate individual expected benefits theories, we begin with the formal IEB model by \citet{resnick_starting_2012}. \citeauthor{resnick_starting_2012} formalize their logic in equations but but do not elaborate rules with enough detail to define behavior for the computational agents in our simulation. As a result, we first need to translate general concepts from the \citeauthor{resnick_starting_2012} model into rules specific enough for agents to follow.

Following \citeauthor{resnick_starting_2012}, we model participation decisions as a function of community size and anticipated future growth. For any agent, the expected benefit of joining a community equals participation benefits ($B_P$) plus early adopter benefits ($B_{EA}$) minus startup costs $(C_S)$. \citeauthor{resnick_starting_2012} treated community success as dichotomous, claiming that people benefit from participation in communities that ``succeed'' by growing to a certain (undefined) size but not from participation in those that ``fail.'' We extend their approach by using a continuous representation of participation benefits. The intuition behind doing so is that a community with 1,000 people has more information, opportunities for friendship, and so on, than a community with 100 people, but that the smaller community provides \textit{some} benefits. Support for this choice comes from recent survey work that suggests that small communities provide value to their members \citep{foote_starting_2017}. We expect benefits of size to scale sublinearly so that a community with 1,000 people is not 10 times more valuable than one with 100. Agents therefore calculate participation benefits ($B_p$) as a logarithmic function of their estimate of the future size ($S_f$) of a community:

$$
B_P = log(S_{F} + 1)
$$

In the first set of IEB models, we assume that agents estimate future community size ($S_F$) by observing the current size and age of the community and making linear extrapolations with slope $\frac{S_C}{\mathrm{age}}$ six time steps into the future:

$$
S_F = S_C + 6 \frac{S_C}{\mathrm{age}}
$$

\noindent In this way, early adopter benefits are a function of both the current size of the community, in that joiners of small communities have more opportunities for influence and status, and of the estimated future size ($S_F$), because influence and status are more valuable in larger communities. 

As with benefits from community success, the \citeauthor{resnick_starting_2012} model treats early adopter benefits as dichotomous. In this approach, a subset of early adopters benefit if a community succeeds, while no one else does. As with participation benefits, it makes more sense to model early adopter benefits as continuous so that they increase with the eventual size of the community, but decrease with the size of a community when an agent begins participating. Specifically, agents calculate early adopter benefits ($B_{EA}$) as the ratio of the natural logarithm of the their estimate of the community's future size ($S_F$) and the natural logarithm of the current size if they join ($S_C + 1$). To avoid numeric issues, we add $1$ to $S_F$ and 2 $S_C$:

$$
B_{EA} = \frac{log(S_F + 1)}{log(S_C + 2)}
$$

\noindent Treating benefits as a continuous function makes it possible for an agent to choose the top ranked of the possible communities to join. 

Finally, for the sake of parsimony, we assume that startup costs ($C_s$) are fixed and identical across communities, but only apply to communities an agent does not already belong to. The total individual expected benefits for some agent $i$ for a given community (denoted with the subscript $j$) is therefore:

 $$B_{ij} = B_{Pij} + B_{EAij} - \mathbf{1}_{i \notin j}C_{S}$$

\noindent Where $B_P{ij}$ is the agent's estimate of the participation benefits for community $j$, $B_{EAij}$ is agent $i$'s estimate of early adopter benefits for $j$, and $ \mathbf{1}_{i \notin j}$ indicates when $i$ does not already belong to $j$ (and thus has to pay the startup cost to join).

\begin{figure}
    \centering
    \includegraphics[width=\mywidth]{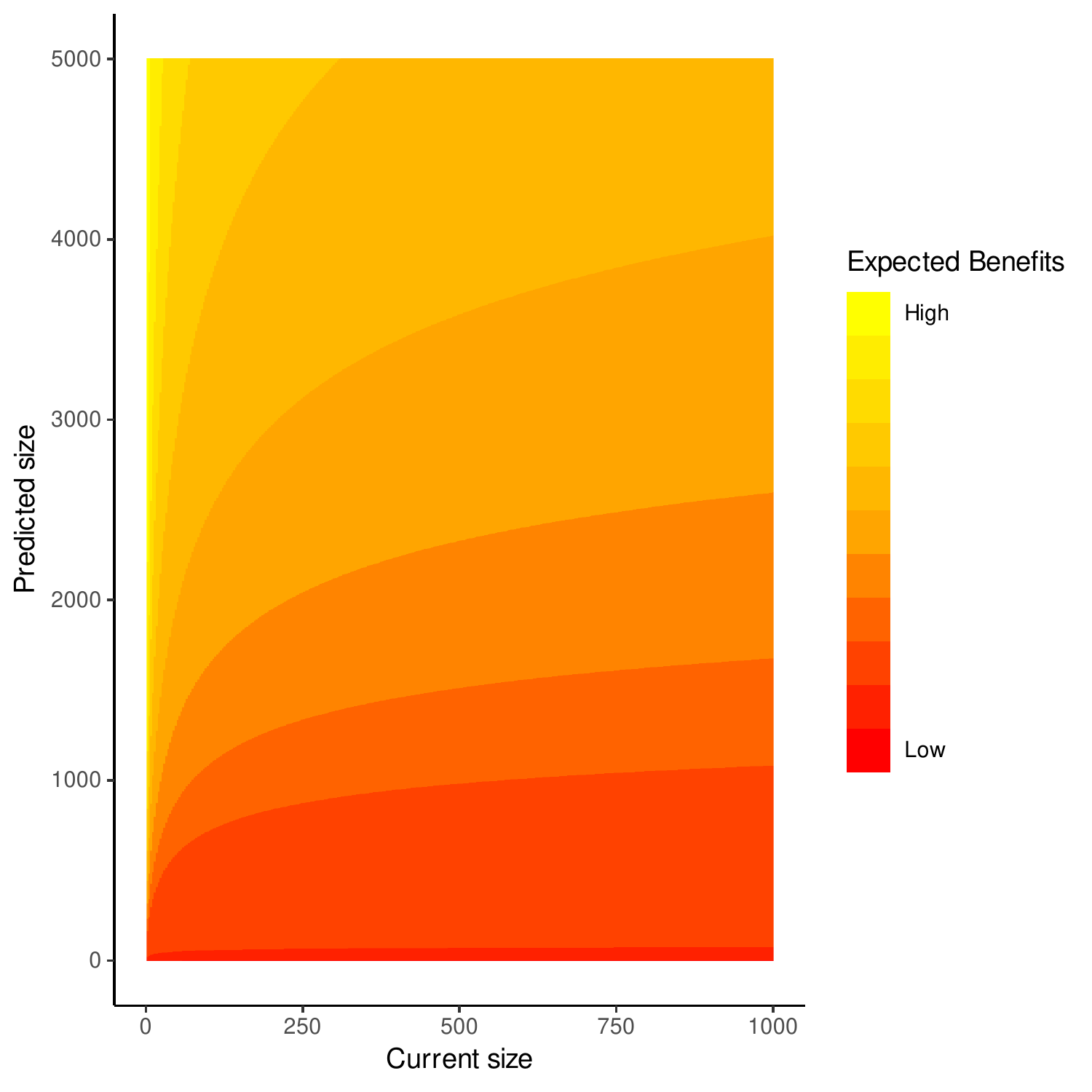}
    \caption{Visualization of total expected benefits in our formalized IEB process. The $x$-axis shows the current size ($S_C$) of the community when the agent considers joining and the $y$-axis represents the agent's prediction for how large the community will grow six time steps in the future ($S_F$)}
    \label{fig:utility}
\end{figure}

Our formalization of the IEB theory captures the main ideas of the \citeauthor{resnick_starting_2012} model and makes more realistic assumptions in several respects. Figure \ref{fig:utility} uses the total expected benefits function to visualize expected benefits over a range of values for the community size when an agent joins it (increasing along the $x$-axis) and the predicted community size (increasing along the $y$-axis). The greatest expected benefits occur in the top-left of the figure representing communities when they are predicted to grow large and when the agent has the opportunity to join early.

As our simulations proceed, agents also make decisions about whether to stay in communities or exit. More formally, an agent may join or exit a set of communities comprised of the union of their current communities and the exposure set in each time step. The agent participates in the proportion $p_k$ of these communities with the highest expected benefits ($B_i$). We know from empirical research that participation is rare, but we don't know how rare. We therefore run simulations across the set of $p_k$ values $\{.05,.1,.2\}$.

As with the social exposure models, we also simulate alternative specifications of the IEB models. Profound skew in community size suggests that people may value community size more highly than linear model terms can accommodate. To address this, we also simulate agents who extrapolate from the current community size using a quadratic function. This increases the predicted size of already large communities by much more than it increases the predicted size of small communities and further reinforces the strong preference for larger communities.

\subsubsection{Combined models}

Finally, we simulate a model that includes both social exposure and IEB decisions. For this combined model, we use the versions of social exposure and IEB intended to produced the most skewed distributions: social exposure when sharing the largest communities and joining and exit determined by the IEB formulas when using a quadratic projection of community size. 

\subsection{Empirical validation}

The sizes of subreddits on Reddit provide the empirical baseline for our comparison. In order to construct a sample that matches the scale of our simulations, we used data published by Pushshift to identify the number of active members of all 23,663 subreddit communities active in January 2017. We define an active member as a unique username that has commented at least five times during the period of data collection. Because the communities in our sample can only have up to 9,000 members, we truncate the plots at that point. This excludes the size data of 29 subreddits which had more than 9,000 members.

We analyze the results of our agent-based simulations through visual comparison of the distributions of community size generated by the simulations and those observed in a sample of online communities. 
We present complementary empirical cumulative distribution functions (eCDFs) to visualize data from our simulations and from real communities. The community size is shown along the $x$-axis. The proportion of communities at least as large as any given value along the $x$-axis are shown along the $y$-axis. Due to the skew of the data, we log both axes. We elaborate on the empirical sample, eCDFs, and the rationale for our comparisons below. 

We inspect our plots to evaluate whether our simulations generate heavy tails reflecting the large number of very large communities found empirically. Such distributions appear as a straight or nearly straight diagonal line on our plots. In contrast, normal (Gaussian) distributions deviate quickly and sharply away from the diagonal. Our analysis considers whether the simulated eCDF generated by the different models produce straight lines as well as how well they align with the eCDF from Reddit.

For the different families of simulations, we present the results as grids of plots. Each cell of these grids corresponds to a single permutation of the possible parameter values described above; the permuted parameter names and values appear above and to the right of each grid. Every plot includes the community size eCDF produced by the corresponding simulation in blue as well as the community size eCDF from our sample of subreddits in red-orange. The $x$-axis and $y$-axis labels for the plots appear along the bottom and to the left of each grid. Each curve starts at the top of the $y$-axis at the point corresponding to the smallest community and reaches 0 at the size of the largest community at the corresponding point on the $x$-axis. The eCDF produced by the subreddit communities is identical in every sub-plot of every figure and differences in appearance results from shifts in the scaling of axes and aspect ratios.

Interpretation of our results rests on identifying qualitative similarities between the eCDF from Reddit and the eCDFs from our simulations. While quantitative procedures and statistical tests for comparing CDFs exist, our goal is not to simulate a process that can generate the precise distributions from Reddit. Rather, our goal is to simulate theoretical models of community joining and evaluate if they can generate a distribution of community sizes with characteristics that are qualitatively similar to those observed in reality. For each simulation, we ask whether community members' collective behavior traces a fairly straight diagonal line across the range of the log-log eCDF plot. This shape means that a considerable number of both large and small communities exist.

\section{Results}

Overall, we find that only a synthesis of social exposure and IEB decisions produce community size distributions that broadly resemble empirically observed patterns of behavior. The null model fails to produce realistic distributions of community sizes. Our simulations of agents using either individual expected benefits-based decisions or social exposure rules alone do better than the null model but produce either too few large communities or too few small communities to reproduce empirical patterns. We explain each set of results in more detail below.

\subsection{Null models}

\begin{figure}
    \centering
    \includegraphics[width=\mywidth]{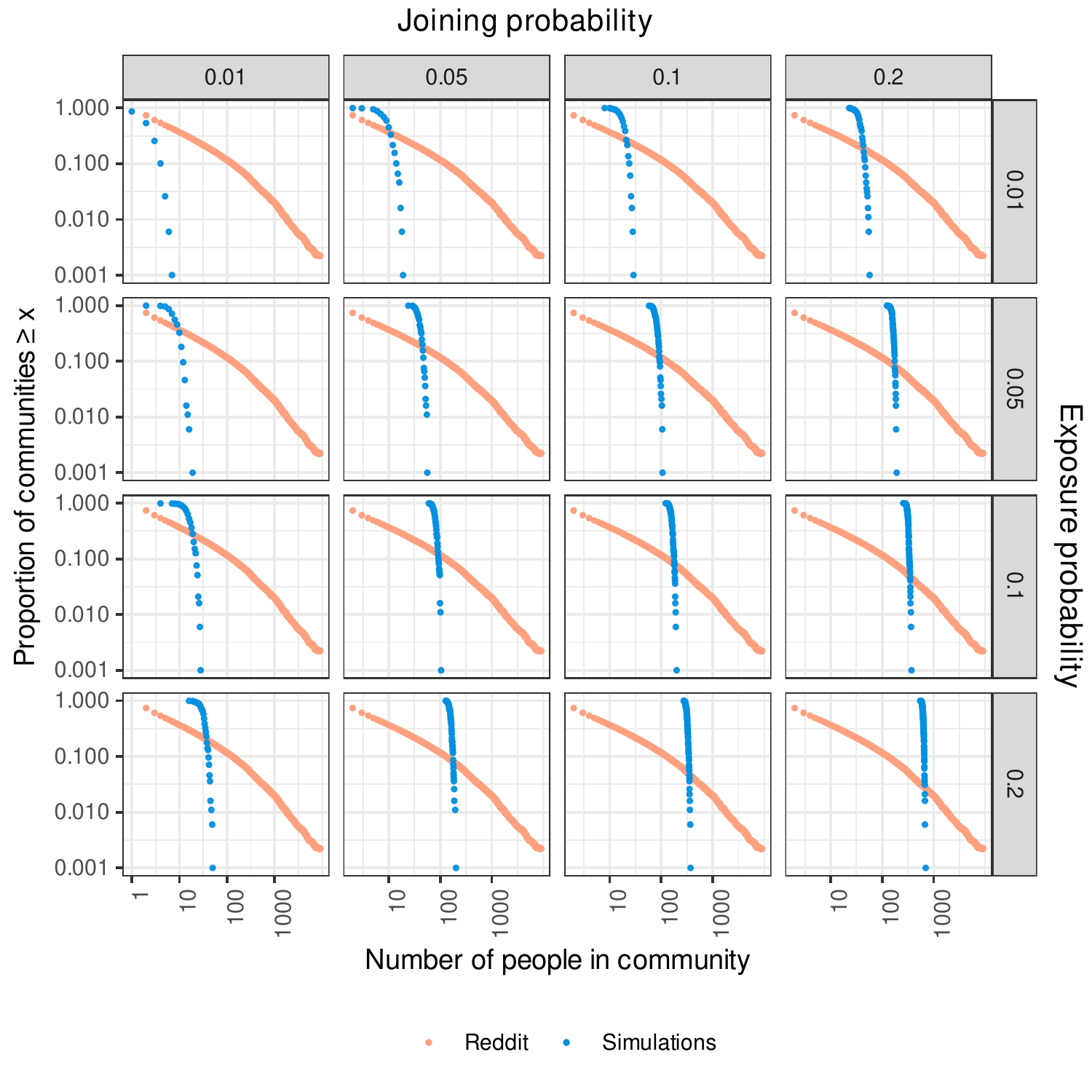}
    \caption{Null models with random exposure and random decisions to join. 
    Moving from from left to right across the grid, each column reflects an increasing probability that agents will join a community they are exposed to.
    Rows reflect increasing probability of exposure from top to bottom.
    Each sub-plot visualizes eCDFs, where the $x$-axis is the proportion of all contributors that are in a given community and the $y$-axis is the number of communities at least that large.}
    \label{fig:comm_null}
\end{figure}

Figure \ref{fig:comm_null} shows a grid of eCDF plots of community sizes for the null model simulations which incorporate random exposure and random joining and exit decisions. The grid of plots show how the simulated eCDF changes with varying joining probability (increasing from left-to-right) and exposure probability (increasing from top-to-bottom).

The simulated eCDFs (in blue) do not produce straight diagonal lines nor do they align with the empirical eCDFs produced by subreddit communities. Starting in the top left of the grid, the simulation with the lowest probabilities of random joining and exit decisions and of random exposure mainly generates small communities. Towards the bottom right of the grid, the joining and exit decision and exposure probabilities get higher and lead to uniformly larger communities. Nearly all of the simulated eCDFs assume a vertical appearance at some point along the community size distribution ($x$-axis). In the context of a log-log CDF plot, this pattern is characteristic of more bell-shaped (normal) distributions with relatively low variance.

When the probability of exposure and joining and exit decisions are low, the distribution is slightly skewed. At higher probabilities, it is bell-shaped. The overall similarity of these results across the grid suggests that none of the parameters in the ranges we include have a strong direct effect on the skew of community sizes.

\subsection{Social exposure models}

\begin{figure}
    \centering
    \includegraphics[width=\mywidth]{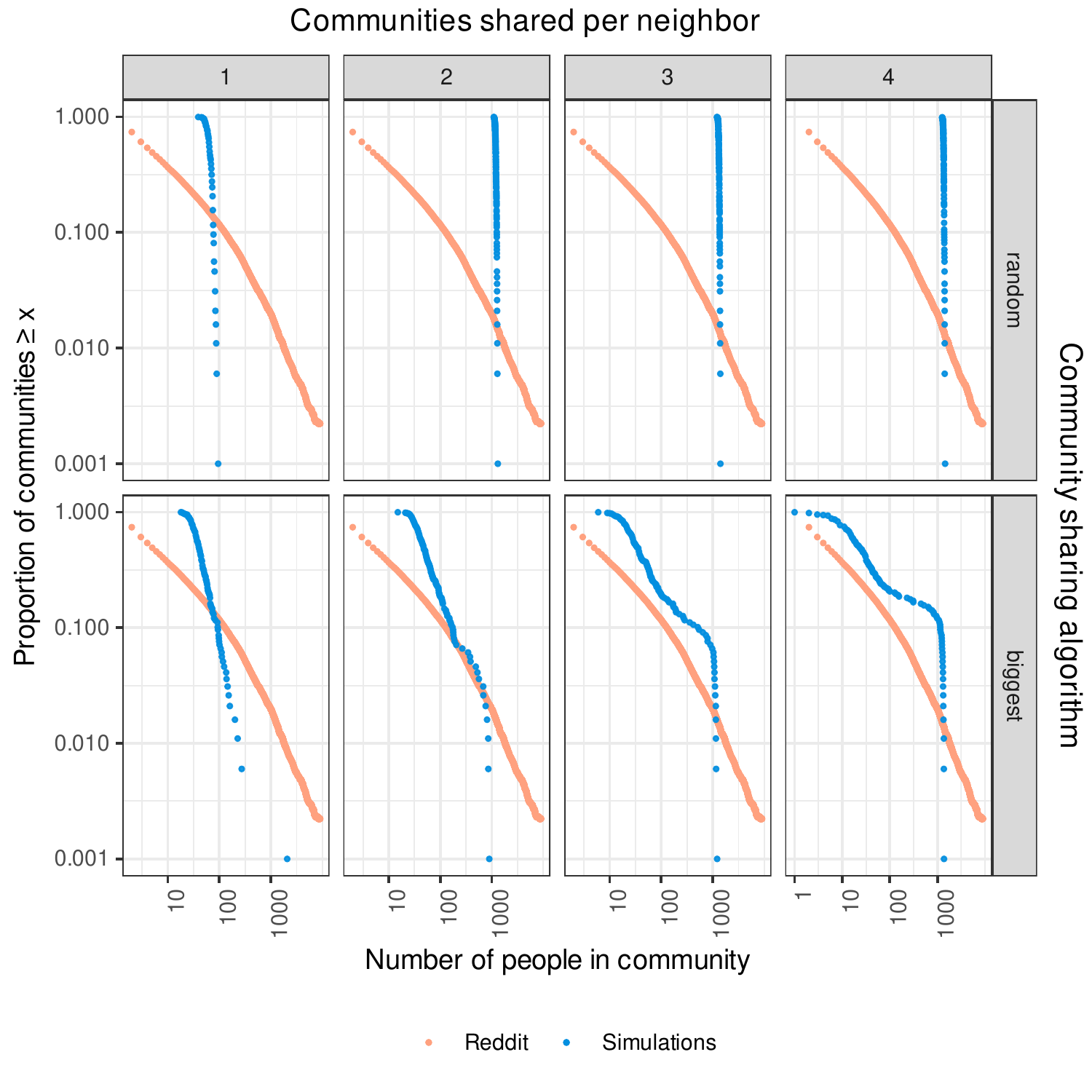}
    \caption{Social exposure models. Complementary eCDF plots showing when agents are exposed to communities via others in their current communities. Moving from left to right, the number of communities that each ``neighbor'' shares increases. Upper plots show when agents share a random set of communities, lower plots show when they share the largest communities to which they belong.}
    \label{fig:social_exp_random}
\end{figure}

Figure \ref{fig:social_exp_random} plots the results from simulations of individual-level social exposure to new communities along with random joining and exit fixed at $.1$ and $.56$, respectively.\footnote{The appendix includes results from a model with joining set at $.05$.} In this grid, the number of communities shared per neighbor increases over the columns from left to right. The two rows contain versions of the models where agents share randomly selected communities (top) and the biggest communities (bottom).
The top row suggests that social exposure when agents share a random set of communities generates a distribution of communities of similar sizes rather than a range of large and small communities. This occurs no matter how many communities each neighbor shares---even as the size of communities grows as the number of shared communities increases.

We find that models in which agents share only the largest communities to which they belong (the bottom row of Figure \ref{fig:social_exp_random}) produce more realistic distributions of community size. As the number of communities shared per neighbor increases (looking left-to-right across the row) the simulated eCDF shifts closer to the diagonal and the eCDF from Reddit. However, substantial deviations remain even in the bottom right plots that offer the best fit. These simulations produce too few small communities and too few large communities to line up more closely with the empirical data. Visually, this is why the simulated eCDFs in this bottom row start out above the eCDF from Reddit and then cross below it at some point along the X-axis.

\subsection{IEB models}

\begin{figure}
    \centering
    \includegraphics[width=\mywidth]{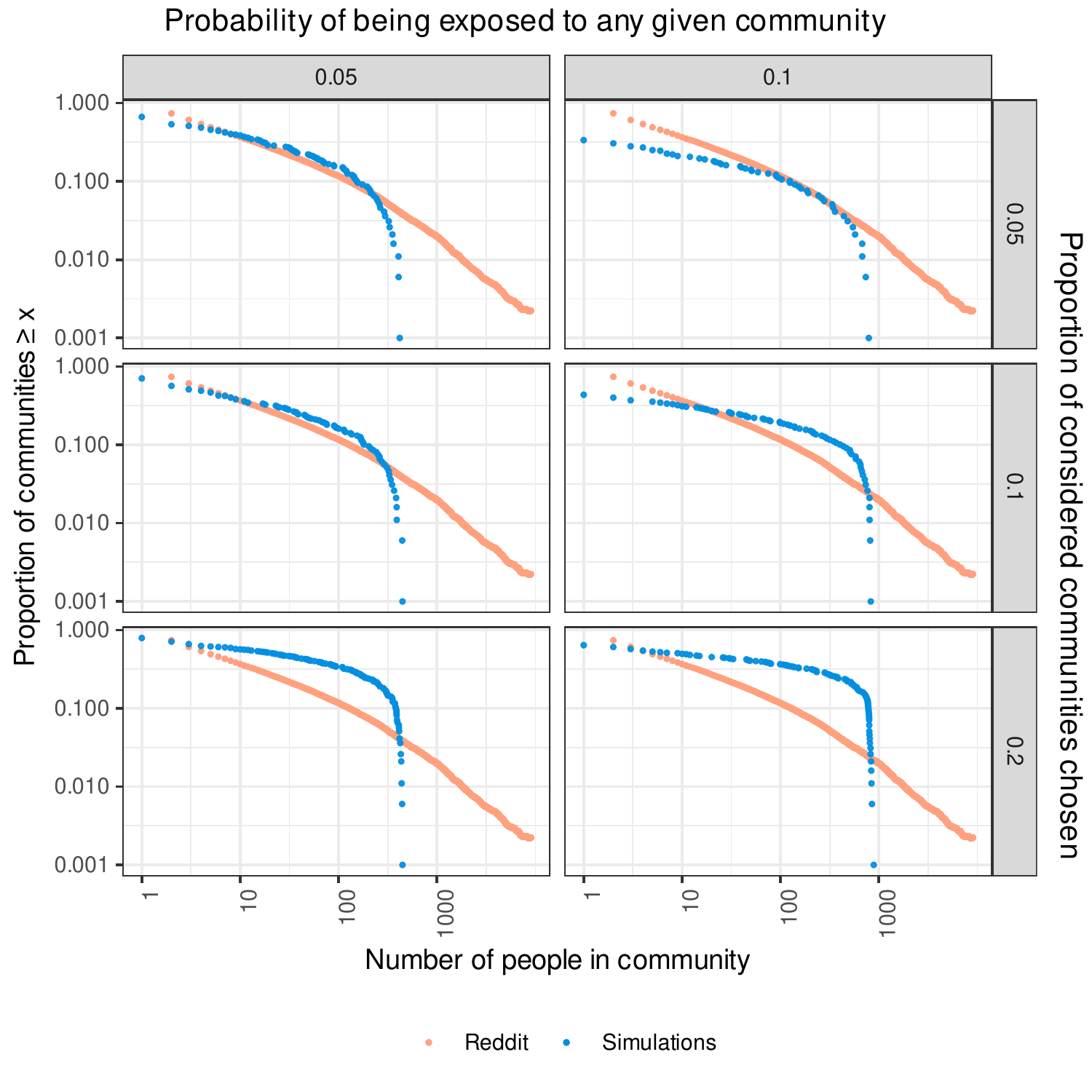}
    \caption{IEB models. Complementary eCDFs of community size
    when agents are exposed to a random set of communities and choose based on IEB. Moving from left to right, the proportion of communities that agents join increases. From top to bottom, the probability of random exposure increases.}
    \label{fig:random_linear}
\end{figure}

Figure \ref{fig:random_linear} shows results of simulations with agents that make joining and exit decisions based on individual expected benefits after being exposed to a random subset of communities. The rows of the grid vary the proportion of communities from the exposure set that each agent joins. The columns vary the value of the random exposure probability ($p_e$). 
As with the null models and the social exposure models, none of the IEB models produce results that align with the empirical baseline across the full range of the eCDFs.

The IEB models generate better fits to the empirical data than either the null models or social exposure models. The best fits in the grid appear along the left column and towards the top row. This suggests that agents making IEB decisions and joining a small proportion of the largest communities they are randomly exposed to produces the most realistic distribution of community sizes. However, the simulated results remain consistently less skewed than the empirical data at the high end of the community size range. The single best fitting simulated CDF in Figure \ref{fig:random_linear} (in the top-left) tracks the eCDF very closely across most of the community size range, and then falls off. This illustrates that the IEB decision models also fail to generate a sufficient number of the very largest communities. The second IEB model, in which agents fit a quadratic rather than a linear equation, produced nearly identical results. A figure showing these results is included in the Appendix.

\subsection{Combined models: Social exposure plus IEB}

\begin{figure}
    \centering
    \includegraphics[width=\mywidth]{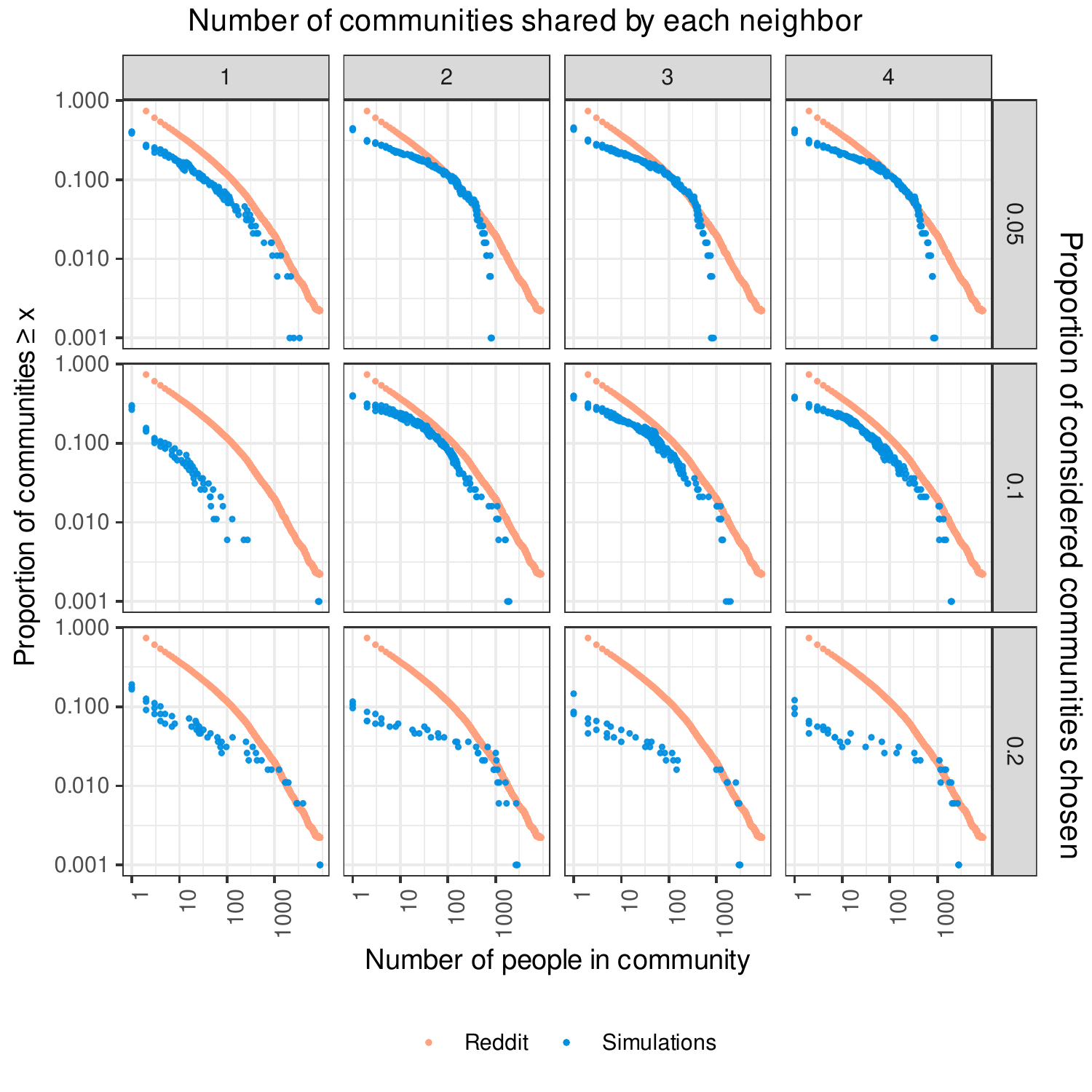}
    \caption{Combined model. Community sizes when agents are exposed to new communities via social exposure and  make participation decisions based on IEB. Moving from left to right, the number of communities that each  ``neighbor'' shares increases. Moving from top to bottom, the proportion of the considered communities that an agent chooses to join or remain in increases.}
    \label{fig:comm_combined}
\end{figure}

The combined models incorporate the two sub-processes of community joining and exit considered separately above. Figure \ref{fig:comm_combined} shows the results of these models across a range of parameters for each of the sub-processes. In terms of social exposure, the plots in the grid show agents that share an increasing number of communities with their neighbors, from left to right. In terms of IEB decisions, the plots display agents that join an increasing proportion of communities from their exposure set, from top to bottom.

As before, none of the plots in this grid show perfect alignment between the simulated and empirical data across the full range of both the $x$ and $y$ axes. In general, the deviations once again emerge at the upper end of the community size distribution, indicating that even both social exposure and IEB decisions combined fail to generate a sufficient number of the very largest communities within the framework of our simulations. However, several of these plots align much more closely than any of the others we have presented thus far and many of them appear as straight or nearly straight lines. In particular, the simulation data plotted in the top left cell and the plots in the right three cells of the middle row line up closely with the eCDF right up until the very upper end of the community size distribution. Other cells along the top row also show good alignment.

The additive effects of the two sub-processes likely explains the improved fit we observe among several of the combined models. The fact that the improved fit recurs across multiple parameter values suggests that the improvements come from the features of the model rather than any particular parameter value on its own. In the appendix, we show the results of a robustness test with lower exposure probabilities and find similar results. 

\section{Discussion}

Our simulations suggest that the combination of social exposure and IEB---two mechanisms of community joining and exit identified in previous research---produce realistic distributions of community size. Simulations based on social exposure or IEB alone result in right-skewed community sizes but do not generate extremely large communities. A combined model produces community sizes that more closely resemble the entire range of empirical community sizes. These findings extend prior research by providing evidence of this joint relationship. The ABS results also illuminate aspects of the relationships between the levels of individual and collective behavior that prior research had not considered.

A few findings bear further comment. First, when we model agents exposed randomly to communities, no decision process produces enough really large communities. Random exposure cannot explain empirical distributions of community size in the absence of a plausible mechanism of exposure to a much larger proportion of communities. Mechanisms that make exposure to larger communities more likely than exposure to smaller ones are required to produce extremely large communities. Social exposure provides a compelling mechanism that does this and has prior empirical and theoretical support.

However, even the effect of social exposure depends on which communities agents share. When agents share communities randomly, social exposure also has little impact on the distribution of community sizes. The first set of social exposure simulations suggest that when people belong to only a few communities and share randomly, no community gets large enough to benefit from the feedback mechanism introduced by social exposure. Our second set of social exposure simulations uses information about patterns in community size to modify our initial model so that agents choose to share the largest communities to which they belong. Only then does social exposure lead to skewed community sizes. This suggests an extension to our understanding of social exposure: that people are more likely to share large communities. We believe this reflects a novel proposition that should be tested empirically.

That said, this mechanism appears to have some limits. Even when agents only share the largest communities they belong to, community sizes remain concentrated and very few communities reach the massive sizes we observe in Reddit. Figure \ref{fig:social_exp_appendix} in our appendix shows additional analyses that suggest that this shortcoming appears even more pronounced when agents are initially exposed to a smaller proportion of the community space.

Another initially perplexing finding provides a clue as to what may be happening. Although we might expect the most extreme skew when agents share only a few of the very largest communities they belong to (thus making them ever larger), the simulations suggest the greatest skew emerges when agents share many. If we look closely at Figure \ref{fig:social_exp_random} we see that large communities are always produced and what improves the fit is the production of more moderately large and small communities. One explanation is that by only sharing one community, people quickly converge on one or two large communities. When more communities are shared, this includes more medium-sized communities.
% One explanation is that limited sharing may lead people to belong to small clusters so that they are simply sharing the same overlapping communities with each other. By sharing multiple large communities, more agents may gain exposure to each other and to a broader range of communities. Future research could explore directly the emergence of clusters from similar models and compare these to participation clusters in online community platforms.

Our finding that neither social exposure nor IEB decisions alone could produce the extent of cumulative advantage necessary to generate the very largest communities also extends prior work. On their own, each sub-process provides a plausible social mechanism for cumulative advantage. Previous studies analyzing social exposure or IEB decisions separately had neither evaluated directly what kinds of macro-level outcomes they produce nor considered the implications of the two processes interacting. Similarly, previous studies that identified cumulative advantage as a mathematical mechanism of observed community size distributions had not evaluated whether or to what degree specific micro-level behaviors could approximate inequality in the distribution of individuals across communities. The ABS results we present bridge this divide and advance both bodies of prior work simultaneously. 

Our findings are far from obvious and the simulations presented might have produced very different results. For example, at the outset of this project, we did not know whether both sub-processes would be necessary. Nor did we know whether agents who share and join the largest communities would create a bi-modal distribution where nearly all communities were small and all agents belonged to a few very large communities. Just because we, the researchers, designed and manipulated the parameters of the simulated models does not mean we could anticipate or determine the results. 

The reasons that social exposure and IEB decisions produce a good approximation of the full distribution deserves attention in future research. For example, we expect that clustering may play an important role in explaining why agents who preferentially share and join the largest communities nevertheless wind up in many small and medium sized communities. Such a pattern might occur because many communities remain "unheard of" outside of local networks of agents who share overlapping communities. This proposition merits further evaluation and might help explain other dimensions of empirically observed behavior.

In sum, our simulations suggest that social exposure combined with IEB decision-making provide a reasonable explanation for empirical community size patterns. Our results also indicate that there is more to the story, however, and that these processes only explain much of the skew in community size. Below we discuss other features that future work should consider.

\subsection{Limitations}

Our approach has several important limitations. The most important is common to all agent-based simulations. While we chose a set of models that we believe capture the most important aspects of the real-world social computing system we seek to understand, other reasonable formulations might lead to different outcomes. Guided by theory, we attempted to identify the aspects of the model most likely to be key sources of variation and to parameterize and test those aspects. Although we are confident that our models are useful and valid in their current form, we have no doubts that they can be productively elaborated upon.

ABS, like all scientific modeling, intentionally elides details of the real world systems we seek to understand. For example, we follow most other ABSs in treating all of the agents and collectives in our simulations as homogeneous. While our model provides plenty of opportunity for certain types of heterogeneity to arise, it is obvious that real people are heterogeneous in their resources, interests, and skills. Similarly, communities have topics which may be of broad or very narrow interest. Although we chose a more parsimonious approach to modeling that ignores it, we are confident that this heterogeneity contributes to heterogeneity in community sizes and can explain some of the remaining skew in participation our models fail to capture. Other aspects of these systems are also likely to influence group sizes and are deserving of attention, such as the role of pseudonymity/anonymity, heterogeneity in costs to contribute, tools for social interaction, or technological features like recommendation systems or default community memberships.

Finally, our work is limited in that we evaluated our ABSs using only at a single macro-level behavior (community size) from a single empirical data source (Reddit). Future work should look at how well these simulations predict additional outcomes and behavioral patterns at meso- and macro-levels  such as clustering in participation networks, heterogeneity in individual participation rates, or temporal patterns of contribution. Additional studies might also validate or revisit findings against multiple empirical baselines to ensure that conclusions do not reflect the biases of a single platform, interface, or time period.

\section{Conclusion}

Two social computing theories of how individuals learn about, join, and leave communities provide a reasonable explanation for how highly unequal distributions of community sizes arise. These results link micro-level models of social exposure and IEB joining and leaving decisions to a distinct scholarship on population-level distributions of community size. The results also support novel theoretical extensions of these micro-level models that can be tested empirically.

In practical terms, the results underscore that highly skewed and unequal community sizes need not result from failures on the part of community leaders or participants. Instead, macro-level inequalities likely arise through the aggregation of individual tendencies and preferences magnified by the massive scale of large sites like Reddit and others. The fact that two very simple processes explain so much about community sizes also suggests that platforms should be cautious when changing things that could directly impact either what people share or the information visible to help them make decisions. Designers, community managers, advertisers and others may want to nudge users towards broader and more equitable community sizes. However, doing so may require fundamentally transforming the ways that individuals learn about or decide to participate in their communities.  

Micro-macro divides such as the one explored in this study provide many opportunities for future social computing research. Theoretically-grounded ABS combined with empirical validation provides an ideally-suited approach to advancing these inquiries. We hope others will extend and evaluate the results presented here, both directly through modeling other macro-level aspects of online communities, and through using agent-based simulations to bridge other micro-macro divides to contribute to our understanding of social computing systems.

\begin{acks}
An earlier version of this paper was part of the first author's PhD dissertation, and some of the text and images appear in that thesis. This work was supported by NSF grants IIS-1617468, IIS-1617129, IIS-1908850 and IIS-1910202. Versions of this paper received very helpful feedback from participants at the International Communication Association conference, the Organizational Communication Mini Conference, and the International Conference on Computational Social Science. The simulations were run on the University of Washington's Hyak computing cluster.
\end{acks}

\bibliographystyle{ACM-Reference-Format}
\bibliography{refs}

\newpage
\appendix
\section*{Appendix A: IEB model with quadratic projection}

\begin{figure}[htb!]
    \centering
    \includegraphics[width=\appwidth]{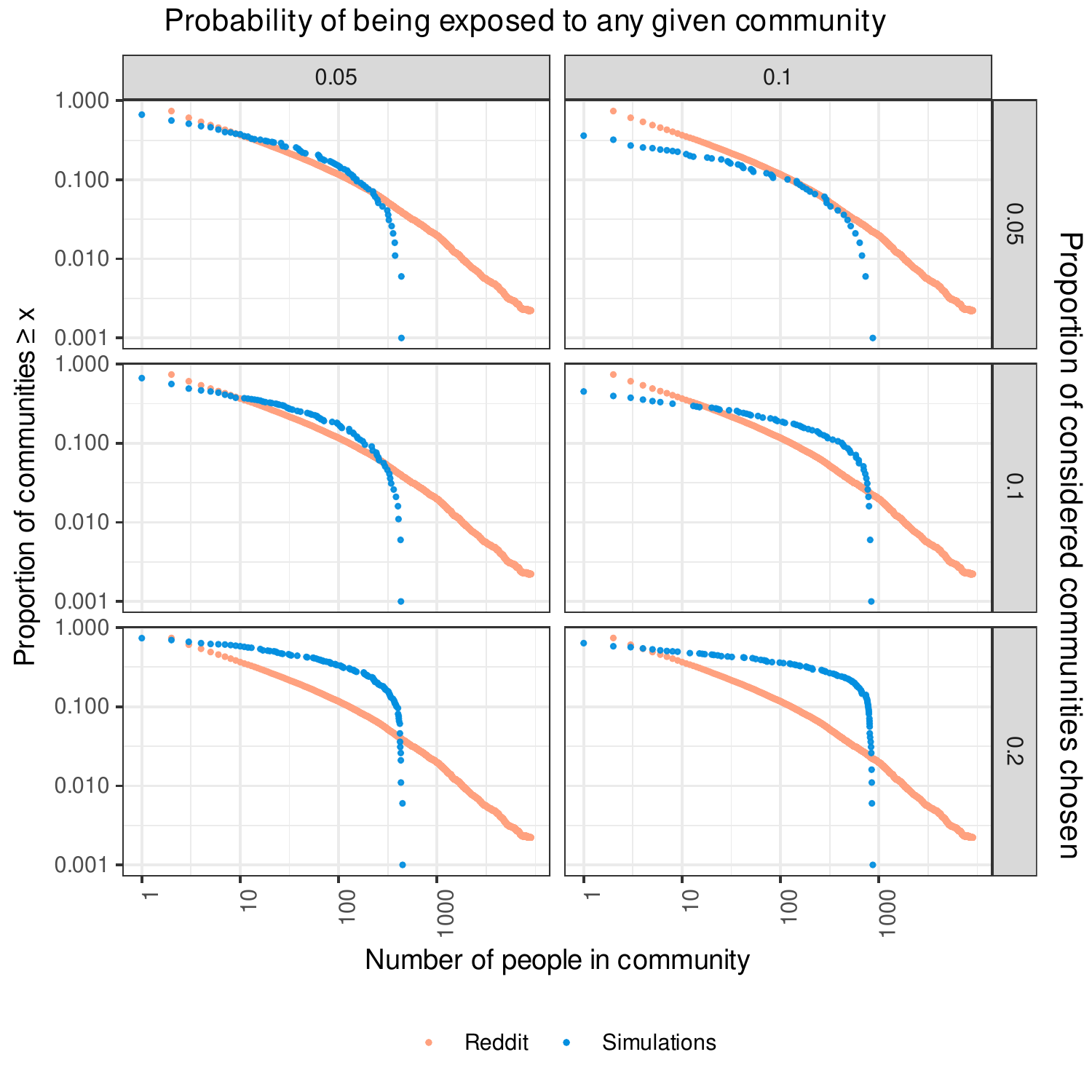}
    \caption{IEB model with quadratic projection. Results when agents are exposed to a random set of communities and use IEB to decide which to join or leave. In these simulations, agents use a quadratic equation to estimate future community size.}
    \label{fig:random_linear_squared}
\end{figure}

Figure \ref{fig:random_linear_squared} shows the results of a set of simulations where exposure is random and joining and leaving decisions are made based on the IEB equations. The results are nearly identical to Figure \ref{fig:random_linear}, suggesting that people were already joining the largest projects.

\section*{Appendix B: Robustness check results}

\subsection*{B.1 Social Exposure and Random Joining}
\begin{figure}[htb!]
    \centering
    \includegraphics[width=\appwidth]{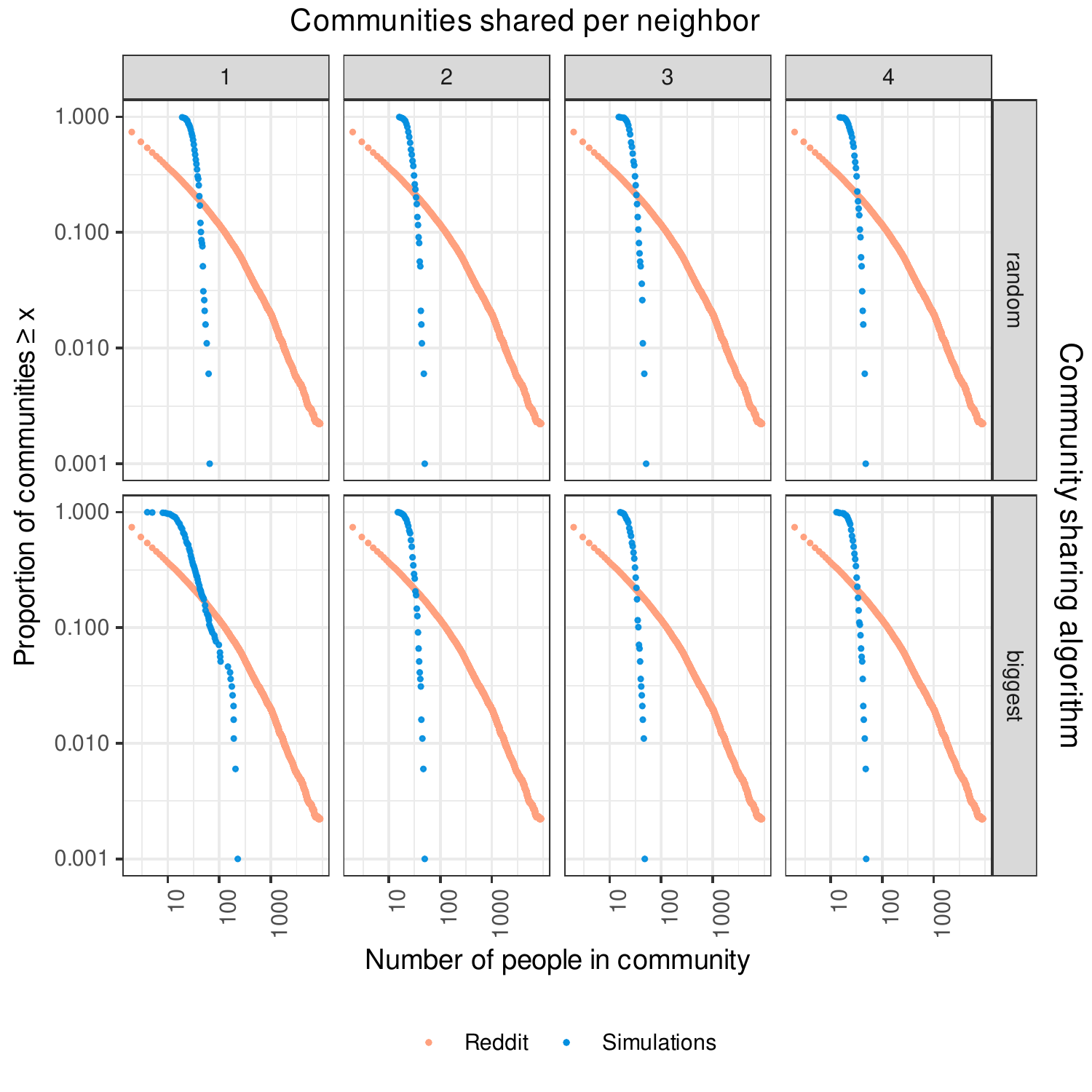}
    \caption{Community sizes
    when people are exposed to new communities via people in their current communities sharing the largest communities to which they belong. Moving from left to right, the number of communities that each ``neighbor'' shares increases. On the top neighbors share random communities. On the bottom they share the biggest to which they belong.}
    \label{fig:social_exp_appendix}
\end{figure}

Figure \ref{fig:social_exp_appendix} shows that when people are less likely to join new communities, nearly all of the skew of community size disappears, suggesting that the influence of social exposure by itself is fragile.

\subsection*{B.2 Combined Model}

\begin{figure}
    \centering
    \includegraphics[width=\appwidth]{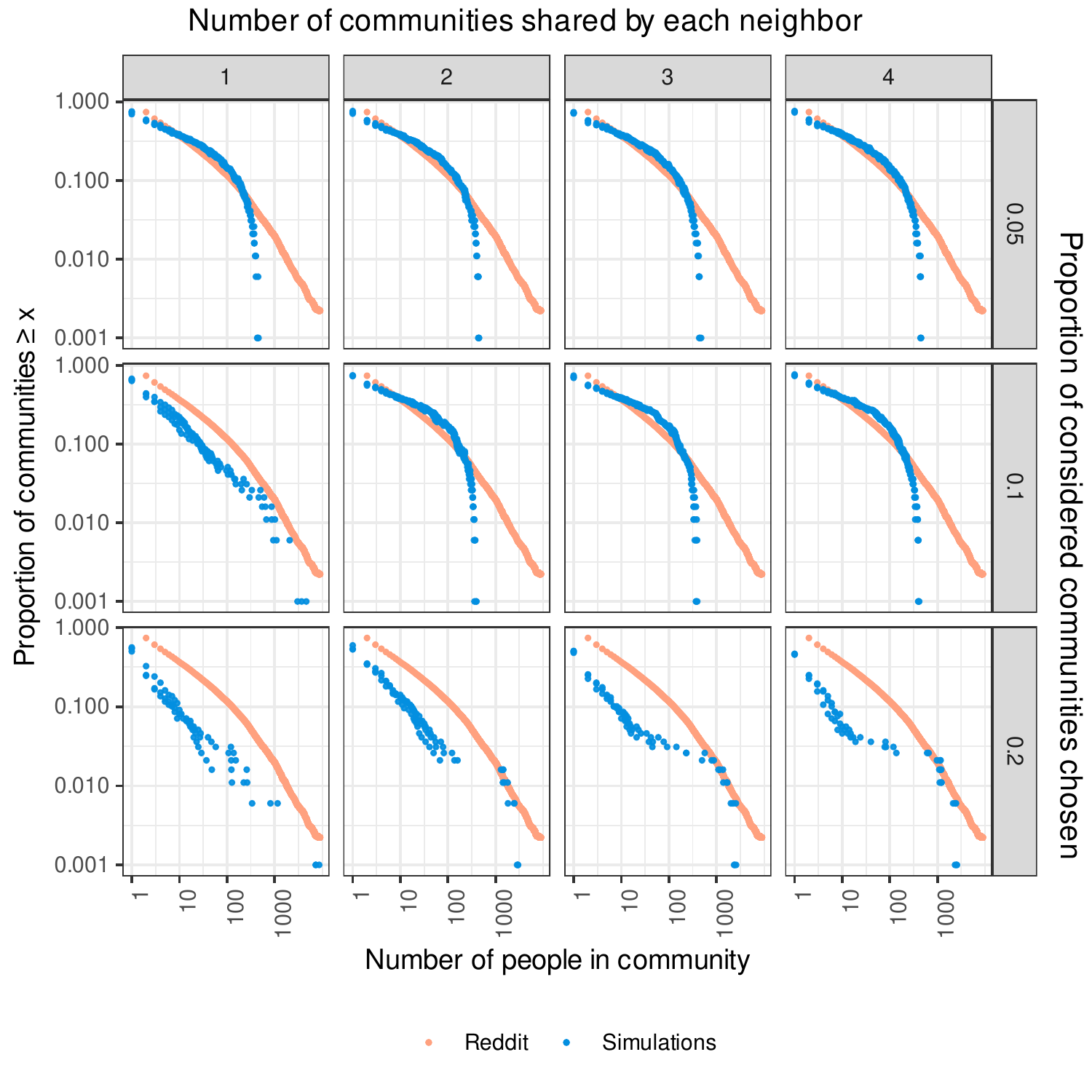}
    \caption{Combined model with the initial probability of exposure set to $.05$. Community sizes when agents are exposed to new communities via social exposure and  make participation decisions based on IEB. Moving from left to right, the number of communities that each  ``neighbor'' shares increases. Moving from top to bottom, the proportion of the considered communities that an agent chooses to join or remain in increases.}
    \label{fig:comm_combined_appendix}
\end{figure}

On the other hand, Figure \ref{fig:comm_combined_appendix} shows that varying the initial proportion of communities that people are exposed to has very little effect on the overall shape of the eventual outcomes.

\end{document}